\documentclass[fleqn,twoside]{article}
\usepackage{espcrc2}

\usepackage{epsfig}
\usepackage[figuresright]{rotating}

\newcommand{\AmS}{{\protect\the\textfont2
  A\kern-.1667em\lower.5ex\hbox{M}\kern-.125emS}}
\newcommand{\bes}{\begin{eqnarray}}
\newcommand{\ees}{\end{eqnarray}}
\newcommand{\bea}{\begin{array}}
\newcommand{\ea}{\end{array}}
\def\Dw{D_{\rm w}}
\def\npf{n_{\rm pf}}
\newcommand{\eq}[1]{eq.~(\ref{#1})}
\newcommand{\fig}[1]{Fig.~\ref{#1}}
\def\gbar{\bar{g}}
\def\mbar{\overline{m}}
\def\fm{\,{\rm fm}}
\def\dtau{\delta\tau}
\def\ZP{Z_{\rm P}}
\def\tauint{\tau_{\rm int}}
\newcommand{\ev}[1]{\langle #1 \rangle}
\def\sigp{\sigma_{\rm P}}
\def\Lmax{L_{\rm max}}

\title{
{
\vspace{-3.0cm} \normalsize \hfill
\parbox{30mm}{HU-EP-03/61\\DESY 03-137\\SFB/CPP-03-38\\September 2003}
}\\[15mm]
       Simulating the Schr{\"o}dinger functional with two pseudo-fermions:
       algorithmic study and the running mass
       \thanks{Talk presented by F. Knechtli.}
       }

\author{F. Knechtli\address[HU]{Institut f{\"u}r Physik, 
        Humboldt Universit{\"a}t, Newtonstr. 15, 12489 Berlin, Germany},
        M. Della Morte\address[DESY]{DESY Zeuthen, Platanenallee 6, 15738 Zeuthen, Germany},
        J. Rolf\addressmark[HU], R. Sommer\addressmark[DESY], 
        I. Wetzorke\address[DNIC]{NIC/DESY-Zeuthen, Platanenallee 6, 15738 Zeuthen, Germany},
        U. Wolff\addressmark[HU] (ALPHA collaboration)
       }

\begin{document}

\begin{abstract}
We present an algorithmic study for the simulation of two massless flavors of
O(a) improved Wilson quarks with Schr{\"o}dinger functional boundary conditions.
The algorithm used is Hybrid Monte Carlo with two pseudo-fermion
fields as proposed by M. Hasenbusch. A gain in CPU cost of a factor two
is reached when compared to one pseudo-fermion field
due to the larger possible step-size. This study is integrated in the ALPHA
project for the computation of the running of the renormalized quark mass.
We include an update on these physics results.
\vspace{-0.5cm}
\end{abstract}

\maketitle

\section{HMC with two pseudo-fermion fields}
\vspace{-0.1cm}
We study \cite{alphalgo} a variant of the Hybrid Monte Carlo (HMC) algorithm
that uses two pseudo-fermion fields per degenerate flavor doublet, as proposed
by M. Hasenbusch \cite{MH1,MH2} and recently tested in \cite{QCDSF}.
The Wilson-Dirac operator with Schr{\"o}dinger functional (SF) boundary conditions
$\Dw+m_0$ is considered together with O($a$) improvement $\delta D$ respectively $T$
(clover term) and {\em e}ven-{\em o}dd preconditioning
\bes
&& a(\Dw + \delta D + m_0) = \frac{1}{2\kappa}M \,,\\
&& M = \left( 
\begin{array}{cc} 1+T_{ee} & M_{eo} \\ M_{oe} & 1+T_{oo}
\end{array}
\right) \,.
\ees
For one mass-degenerate flavor doublet of quarks the partition function reads
\bes
 Z & = & \int_U\exp(-S_{\rm g}(U))\det(1+T_{ee})^2\det{\hat{Q}}^2 \,,\label{znf2} \\
 \hat{Q} & = & \tilde{c}_0\gamma_5\{1+T_{oo}-M_{oe}(1+T_{ee})^{-1}M_{eo}\}\,,
\ees
with $\tilde{c}_0=(1+64\kappa^2)^{-1}$ and $S_{\rm g}(U)$ is the Wilson plaquette
action. To simulate \eq{znf2} we use the Hybrid Monte Carlo algorithm.
The determinant $\det{\hat{Q}}^2$ is represented in terms of pseudo-fermion fields $\phi$.
If $\hat{Q}$ is factorized in $\npf$ factors then one pseudo-fermion field for each
factor is introduced. The aim of the factorization is to make the fermionic contribution
to the forces associated with each factor as small as possible.
We take $\npf=2$ and the factorization
\bes
&& \hat{Q} = (\tilde{Q})(\tilde{Q}^{-1}\hat{Q}) \,, \quad
\tilde{Q} = \hat{Q}-i\rho \,,
\ees
which leads to the pseudo-fermion actions $i=1,2$
\bes
&& S_{{\rm F}_i} =
 \phi_i^{\dagger}\left[\sigma_i^2+(\hat{Q}^2+\rho_i^2)^{-1}\right]\phi_i \,,
 \label{sfi} \\
 && \sigma_1=0\,, \; \rho_1=\rho\,, \quad
 \sigma_2=\frac{1}{\rho}\,, \; \rho_2=0 \,.
\ees
The real parameter $\rho$ is chosen to minimize the sum of the condition
numbers of the two operators appearing in \eq{sfi},
\bes
 \rho & = &  
 \left[\lambda_{\min}(\hat{Q}^2)\lambda_{\max}(\hat{Q}^2)\right]^{1/4} \,,
\ees
in terms of the smallest and largest eigenvalue of $\hat{Q}^2$. The operators
in \eq{sfi} have then both condition number equal to the square root of the
condition number of $\hat{Q}^2$.

This algorithmic study is part of the ALPHA large scale simulations for the
computation of the running of the renormalized quark mass \cite{mbar1}. For
the temporal and spatial lattice size, the background gauge field and the
parameter controlling the spatial boundary conditions of the fermion fields
we take respectively
\bes
&& T=L\,, \quad C=C^{\prime}=0\,, \quad \theta=0.5 \,.
\ees
The theory is simulated along the critical line where the PCAC mass vanishes.
These simulations are possible in the SF due to the infrared cut-off proportional
to $1/T^2$ in the spectrum of the Dirac operator squared. In our simulations
the renormalized coupling $u=\gbar^2(L)$ \cite{coupling}
takes values in the range $u=1.0\ldots5.7$. 
In the SF $u$ is presumably a monotonically growing function of $L$ and
our simulations correspond approximately to the range $L=10^{-2}\ldots1\fm$.
%
\begin{figure}[t]
\vspace{-0.7cm}
\centerline{\epsfig{file=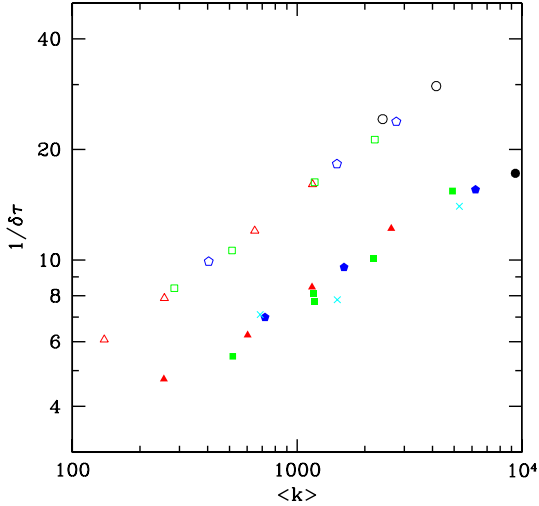,width=8cm}}
\vspace{-1.5cm}
\caption{Step-sizes $\dtau$ for 80\% acceptance. Open symbols are for
$\npf=1$, filled symbols and crosses for $\npf=2$. \label{f_dtau}} 
\vspace{-0.5cm}
\end{figure}
%


\section{PERFORMANCE}
\vspace{-0.1cm}
In the integration of the molecular dynamics equations of motion
we adjust the step-size $\dtau$ to yield an acceptance of 80\% for
trajectories of length one. \fig{f_dtau}
shows the inverse step-size, i.e. the number of steps, as a function of the
average condition number of $\hat{Q}^2$
\bes
  \ev{k} = \Big\langle\frac{\lambda_{\max}(\hat{Q}^2)}
                           {\lambda_{\min}(\hat{Q}^2)}\Big\rangle \,.
\ees
Data are shown for different lattice sizes $L/a$ and renormalized couplings 
$u=1.0,1.1$ (triangles),
$u=2.5$ (squares), $u=3.3$ (pentagons), $u=5.7$ (circles) together with
simulations at $\beta=5.2,\;\kappa=0.1355$ (crosses). The filled
symbols and crosses are obtained with $\npf=2$ and the open symbols with
$\npf=1$ (standard HMC). With $\npf=2$ we can choose step-sizes which are
a factor two larger than with $\npf=1$.
%
\begin{figure}[t]
\vspace{-0.7cm}
\centerline{\epsfig{file=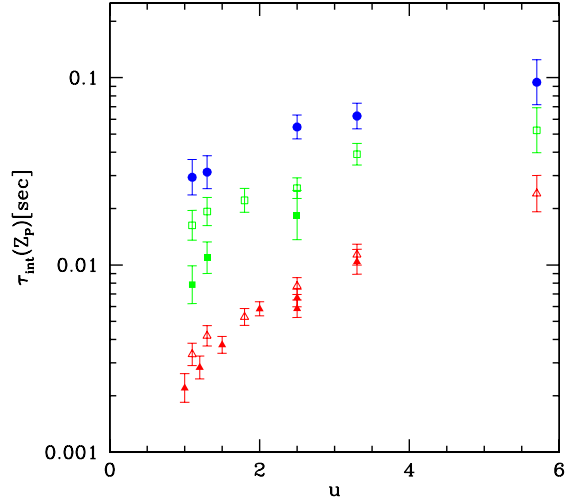,width=8cm}}
\vspace{-1.5cm}
\caption{Integrated autocorrelation time of $\ZP$. Data are for lattices
$L/a=12$ (triangles), $L/a=16$ (squares) and $L/a=24$ (circles), open
symbols for $\npf=1$ and filled symbols for $\npf=2$. \label{f_tauint}} 
\vspace{-0.5cm}
\end{figure}
%

In \fig{f_tauint} we plot
the integrated autocorrelation time $\tauint(\ZP)$ \cite{UWerr}
of the pseudoscalar density $\ZP$ as a function of $u$.
$\ZP$ is the quantity we need in order to compute the
running of the renormalized mass. The units are CPU seconds per lattice point
for the simulation of one lattice on one APEmille crate,
which consists of 128 nodes and has a peak performance of 68 GFlops.
Data are for lattices
$L/a=12$ (triangles), $L/a=16$ (squares) and $L/a=24$ (circles), open
symbols for $\npf=1$ and filled symbols for $\npf=2$.
A systematic difference between $\npf=1$ and $\npf=2$ is seen for the $L/a=16$
lattices, the smallest couplings show a reduction in CPU cost by about a factor
two for $\npf=2$. This CPU gain is due to the larger possible step-size.
A comparison at $L/a=24$ would demand too much CPU time,
precisely because of this it was essential to our project to speed up
the standard HMC.

\section{THE RUNNING MASS}
\vspace{-0.1cm}
%
\begin{figure}[t]
\vspace{-1.8cm}
\centerline{\epsfig{file=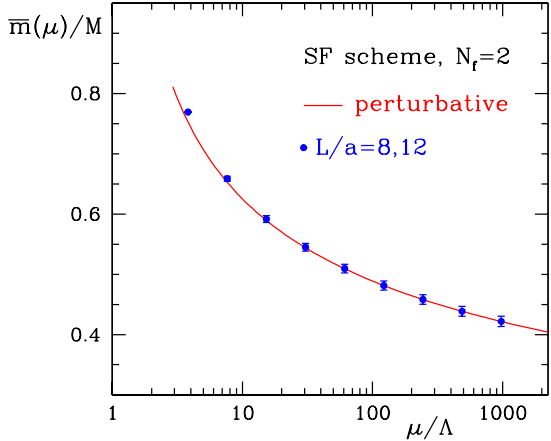,width=9.5cm}}
\vspace{-2.5cm}
\caption{The running of the renormalized quark mass. \label{f_mbar}} 
\vspace{-0.5cm}
\end{figure}
%
The running of the renormalized quark mass $\mbar(\mu)$
with the renormalization scale $\mu=1/L$
is extracted from the step scaling function
$\sigma(u)$ of the coupling and the step scaling function
$\sigp(u)$ of the pseudoscalar density,
\bes
 \sigma(u) & = & \gbar^2(2L)\,, \quad u = \gbar^2(L) \,,\\
 \sigp(u)  & = & 
\left.\lim_{a/L\to0}\,\frac{\ZP(g_0,2L/a)}{\ZP(g_0,L/a)}\right|_{u=\gbar^2(L)} \\
&=& \frac{\mbar(1/L)}{\mbar(1/2L)} \,.
\ees
We solve the joint recursion: $k=0,1,2,\ldots$
\bes
&& \left\{\bea{l} u_0=\gbar^2(\Lmax)=4.6800 \\ \sigma(u_{k+1})=u_k \ea\right. \,,\\
&& \Rightarrow u_k=\gbar^2(L_k) \,,\quad L_k = 2^{-k}\Lmax \,, \label{ucoeff} \\
&& w_0=1\,,\quad w_k=\left[\prod_{i=1}^k\sigp(u_i)\right]^{-1} \,, \label{massrec} \\
&& \Rightarrow  w_k = \frac{\mbar(1/\Lmax)}{\mbar(1/L_k)} \,, \label{wcoeff}
\ees
which gives the running of $\mbar(\mu)$ starting at the scale $1/\Lmax$ defined
by $\gbar^2(\Lmax)=4.6800$.

For the step scaling function $\sigp(u)$ we take the average
$\overline{\Sigma}_{\rm P}^{(8,12)}(u)$ of its values
$\Sigma_{\rm P}(u,a/L)$, which we computed on $L/a=8,12$ (and $2L/a$) lattices
for 6 couplings in the range $u=0.98\dots3.3$. Then we interpolate by the Ansatz
\bes
 \overline{\Sigma}_{\rm P}^{(8,12)}(u) & = & 1 -\ln(2)d_0u + p_2u^2 + p_3u^3
\ees
with $d_0$ perturbative, fitted parameters $p_2$ and $p_3$ and use this fit formula
in the recursion \eq{massrec}.
At high energies contact with the perturbative regime is made
from which the RGI parameters $\Lambda$ and $M$ can be extracted \cite{alphalgo}
\bes
 \ln(\Lambda\Lmax) = -1.34(7) \\
 \frac{M}{\mbar(1/\Lmax)} = 1.30(3) \,.
\ees
\fig{f_mbar} is then obtained from the coefficients \eq{ucoeff} and \eq{wcoeff}. 
The errors of the points come from the statistical errors in the coefficients $w_k$,
the scale ambiguities in the quantities $\Lambda\Lmax$ and $M/\mbar(1/\Lmax)$
are not shown.

The outlook of this work is the determination of $\Lmax$ using available data
from JLQCD \cite{JLQCD} and UKQCD \cite{UKQCD}. A hadronic
scheme has to be set up to determine combinations of $M_u$, $M_d$ and $M_s$.

{\bf Acknowledgement.} We thank NIC/DESY Zeuthen for allocating computer time on
the APEmille machine and the APE group for their support. This work was supported by the 
European Community's Human Potential Programme under contract HPRN-CT-2000-00145
and by the Deutsche Forschungsgemeinschaft in the SFB/TR 09.

\end{document}